*Article*

# Revealing the target electronic structure with under-threshold RABBIITT

Anatoli Kheifets 0000 − 0001 − 8318 − 9408

**Abstract:** The process of reconstruction of attosecond beating by interference of two-photon transitions (RABBITT) reveals the target atom electronic structure when one of the transitions proceeds from below the ionization threshold. Such an under-threshold RABBITT resonates with the target bound states and thus maps faithfully the discrete energy levels and the corresponding oscillator strengths. We demonstrate this sensitivity by considering the Ne atom driven by the combination of the XUV and IR pulses at the fundmanetal laser frequency in the 800 and 1000 nm ranges.

## 1. Introduction

The process of reconstruction of attosecond beating by interference of two-photon transitions (RABBITT) [1,2] has become a widely used tool for attosecond chronoscopy of atoms [3], molecules [4,5] liquids [6] and solids [7,8]. In RABBITT, XUV driven primary ionization is augmented by secondary IR photon absorption or emission. The latter IR-driven processes lead to the same final continuous state whose population depends on the relative phase of the absorption/emission amplitudes. Experimental access to this phase difference makes it possible to convert it to the photoemission time delay and to resolve photoemission on the attosecond time scale.

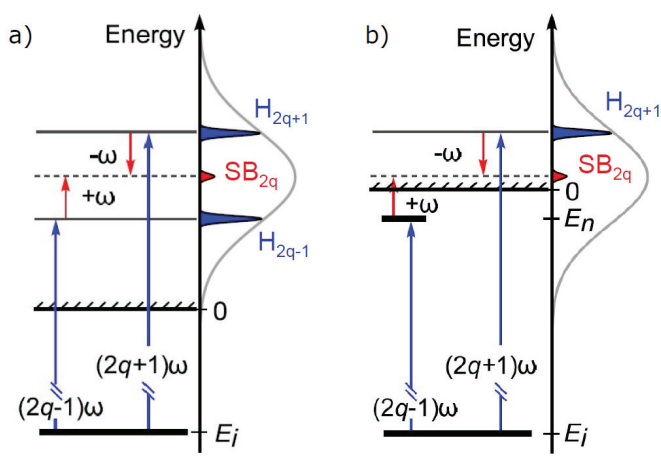

**Figure 1.** a) Schematic representation of the conventional RABBITT process. b) Same for the uRABBITT process.

The RABBITT process is illustrated graphically in Figure 1a). The XUV photon from a high-order harmonics generation source with the frequency $\Omega = (2q \pm 1)\omega$ is absorbed from the initial atomic bound state $E_i$. The ionized electron appears in the continuum and forms the two harmonic peaks in the photoelectron spectrum marked as $H_{2q\pm1}$. The subsequent IR photon absorption $\omega$ or emission $-\omega$ leads to formation of the sideband marked as SB$_{2q}$. Since the two distinct quantum paths lead to the same final state, the SB population oscillates as the time delay $\tau$ between the XUV and IR pulses varies:

$$S_{2q}(\tau) = A + B\cos(2\omega\tau - C).  \quad (1)$$

Here the magnitude $A$ and $B$ parameters depend on the specific experimental conditions. The RABBITT phase parameter

$$C = \Delta\phi_{2q\pm1} + \Delta\phi_W + \Delta\phi_{cc}  \quad (2)$$

is the sum of the phase difference between the neighboring odd harmonics ($\Delta\phi_{2q\pm1} = \phi_{2q+1} - \phi_{2q-1}$), the analogous difference of the Wigner phases ($\Delta\phi_W$) and the difference of the continuum-continuum (CC) phases ($\Delta\phi_{cc}$). The two latter phase differences originate from the XUV and IR photon absorption, respectively. They are converted to the corresponding time delays by the finite difference formula

$$\tau_W = \Delta\phi_W/(2\omega), \quad \tau_{cc} = \Delta\phi_{cc}/(2\omega).  \quad (3)$$

The atomic time delay $\tau_a = \tau_W + \tau_{cc}$ is the group delay of the photoelectron wave packet propagating in the combined field of the ion remainder and the dressing IR field relative to its free space propagation.

The under-threshold RABBITT process, which we term for brevity uRABBITT, is depicted in Figure 1b). In this process, the $(2q-1)\omega$ photon absorption promotes the target electron to a discrete excited state below the threshold $E_n < 0$. It is the subsequent $\omega$ photon absorption that takes the photoelectron to the continuum where it interferes with its downward converted counterpart. The distinct feature of the uRABBITT process is that one the harmonic peaks H$_{2q-1}$ is missing in the photoelectron spectrum. The phase parameter $C$ of the uRABBITT process does not contain the CC component in the IR photon absorption. Instead, it acquires a resonant phase due to the transition between the initial bound to the discrete intermediate states. In the case of an isolated resonance, the resonant phase can be approximated by a simplified expression [9]

$$\Phi_r \approx \arg\left[\Omega + E_i - E_n - i\Gamma\right]^{-1} = \arctan(\Gamma/\Delta).  \quad (4)$$

Here $\Gamma$ is proportional to the spectral width of the XUV pulse and $\Delta \equiv \Omega + E_i - E_n$ is the detuning from the resonance. More elaborate expressions for the resonant two-photon absorption phase are derived in [10,11].

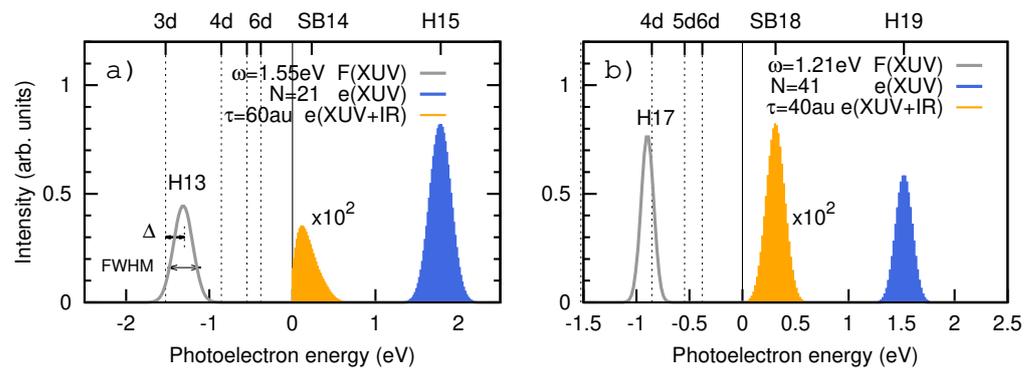

**Figure 2.** a) Simulated photoelectron spectrum of Ne at $\omega = 1.55$ eV (800 nm) is overlapped with the spectrum of the driving XUV pulse. The primary harmonic peak H15 and the scaled sideband SB14 are shaded in blue and orange, respectively. The submerged harmonic peak H13 overlaps with the dicrete 3$d$ energy level. The arrows indicate the detuning $\Delta$ and FWHM. b) Same for $\omega = 1.21$ eV (1024 nm) where the submerged harmonic peak H17 overlaps with the 4$d$ bound state and SB18 emerges above the threshold.

The uRABBITT process has been demonstrated experimentally in He [12] and, more recently, in Ne [13]. The Ne case is illustrated graphically in Figure 2 for the IR laser wavelength near 800 nm (a) and 1024 nm (b). In Figure 2a, the harmonic H15 is matched with the corresponding peak of the photoelectron spectrum while the harmonic H13 falls below the threshold where it overlaps with the 3*d* discrete level. Absorption of an XUV photon from the harmonic H13 leads to the population of the excited 3*d* and 4*s* states with comparable oscillator strengths [14]. However, the IR absorption from 3*d* is much stronger than that from 4*s* and the latter state can be ignored in the uRABBITT process (see Sec. 2.3 for more detail). Similarly, in Figure 2b, the harmonic peak H17 falls below the threshold where it overlaps with the discrete 4*d* level. The uRABBITT sideband SB18 appears just above the threshold.

The RABBITT-uRABBITT phase transition in Ne at 800 nm was studied theoretically [9]. A strong uRABBITT phase variation was demonstrated when SB14 trespassed the 3*d* energy level. A rather large XUV spectral width employed in [9] did not allow for an accurate uRABBITT phase determination in the 1000 nm wavelength range when SB18 was expected to overlap with a group of narrowly spaced target states. In the present work, the XUV spectral resolution is improved and the Ne bound state mapping is reported both in the 800 and 1000 nm spectral ranges. We demonstrate that in both cases the resonant uRABBITT phase maps faithfully the target atom electronic structure thus allowing access to the bound state energies and the corresponding oscillator strengths.

## 2. Theoretical model

### 2.1. Lowest order perturbation theory

The simplest interpretation of the parameters entering Eq. (1) is provided by the lowest order perturbation theory (LOPT):

$$A = |\mathcal{M}_a|^2 + |\mathcal{M}_e|^2 \ , \ B = 2\text{Re}[\mathcal{M}_a \mathcal{M}_e^*] \ , \ C = \arg[\mathcal{M}_a \mathcal{M}_e^*] = 2\omega\tau_a \ . \tag{5}$$

Here we introduce the complex amplitudes of the XUV absorption, augmented by absorption $\mathcal{M}_a$ or emission $\mathcal{M}_e$ of an IR photon. These two-photon ionization amplitudes are written in the LOPT as

$$\mathcal{M}_{a/e} \propto \left\{ \sum_{E_{nl}<0} + \int d^3 k \right\} \left[ \frac{\langle k|d(\omega)|nl\rangle\langle nl|d(\Omega)|i\rangle}{\Omega + E_i - E_{nl} - i\gamma} + \frac{\langle k|d(\omega)|\kappa\rangle\langle \kappa|r|i\rangle}{\Omega + E_i - \kappa^2/2 - i\gamma} \right]$$

Here $\langle i|$, $nl$ or $\langle \kappa|$ and $\langle k|$ are the initial, intermediate and final states, respectively, $d(\Omega)$ and $d(\omega)$ are the dipole operators of the XUV and IR photon absorption. The XUV photon energy is $\Omega = (2q \pm 1)\omega$ and $i\gamma$ denotes the pole bypass in the complex energy plane. For a realistic XUV pulse, the infinitesimal $\gamma$ should be substututed with a finite $\Gamma$ that is proportional to the corresponding FWHM [9].

For the conventional RABBITT, the sum over discrete intermediate states $nl$ is neglected in Eq. (6) as the relevant energy denominators are large. For uRABBITT, the contribution of the discrete sum is essential. We isolate this part of the absorption amplitude into the resonant term

$$\mathcal{M}_{aR} \propto \sum_{nl} A_{nl} \left[ \Omega + E_i - E_{nl} - i\Gamma \right]^{-1} \ , \ A_n = \left[ f_{nl}(\Omega)\sigma_{nl}(\omega) \right]^{1/2}, \tag{6}$$

Here $f_{nl}(\Omega) \propto |\langle nl|d(\Omega)|i\rangle|^2$ is the dipole oscillator strength and $\sigma_{nl}(\omega) \propto |\langle k|d(\Omega)|nl\rangle|^2$ is the partial photo-ionization cross-section. Accordingly, the resonance affected uRABBITT parameters become

$$\begin{aligned} C_r &= \arg\{[\mathcal{M}_a + \mathcal{M}_{ar}]\mathcal{M}_e^*\} = \arg\{[\mathcal{M}_{ar}/\mathcal{M}_a + 1]\} + C \\ B_r &= 2\text{Re}\{[\mathcal{M}_a + \mathcal{M}_{ar}]\mathcal{M}_e^*\} = 2\text{Re}\{\mathcal{M}_{ar}\mathcal{M}_e^*\} + B \end{aligned} \tag{7}$$

The resonance-free *B* and *C* parameters can be extended continuously across the threshold from the non-resonant sideband SB$_{2q+1}$. The unknown analytically amplitudes $\mathcal{M}_{a/e}$ can be found by fitting the numerical TDSE results with Eq. (7).

## 2.2. Non-perturbative treatment

Accurate non-perturbative treatment of the RABBITT process requires numerical solution of the time-dependent Schrödinger equation (TDSE). We seek this solution in the single-active electron (SAE) approximation using the TDSE computer code [15]. The target atom is described by a localized Hartree-Fock potential (LHF) [16]. The TDSE SAE approach to RABBITT has been tested successfully on He [17], Ne [18] and heavier noble gas atoms [19]. The TDSE is driven by a superposition of an XUV attosecond pulse train (APT) and the IR pulse in several fixed increments of the IR/XUV delay $\tau$.

The APT is modeled with the vector potential

$$A_x(t) = \sum_{n=-20}^{20} (-1)^n A_n \exp\left(-2\ln 2 \frac{(t-nT/2)^2}{\tau_x^2}\right) \cos\left[\omega_x(t-nT/2)\right], \quad (8)$$

where

$$A_n = A_0 \exp\left(-2\ln 2 \frac{(nT/2)^2}{\tau_T^2}\right).$$

Here $A_0$ is the vector potential peak value and $T = 2\pi/\omega$ is the period of the IR field. The XUV central frequency is $\omega_x$ and the time constants $\tau_x$, $\tau_T$ are chosen to span a sufficient number of harmonics in the range of photon frequencies of interest for a given atom.

The vector potential of the IR pulse is represented by the cosine squared envelope

$$A(t) = A_0 \cos^2\left(\frac{\pi(t-\tau)}{2\tau_{\text{IR}}}\right) \cos[\omega(t-\tau)]. \quad (9)$$

In the present work, the APT is centered at $\omega_x = 15\omega$ and its spectral width is reduced to $\Gamma = 0.1$ eV by increasing the number of pulselets to $N = 41$ in comparison with $N = 21$ in [9]. Typical XUV and IR field intensities were of the order of $\sim 10^{10}$ W/cm$^2$. In this low intensities regime, our numerical results depend weakly on variation of the pulses intensity. The photoelectron spectrum is obtained by projecting the time-dependent wave function at the end of the time evolution on the basis of Volkov states. Numerical details are given in the preceding publications [18,19]. The present set of calculations required approximately 80K CPU-hours of the Gadi supercomputer hosted at the National Computational Infrastructure (NCI Australia) and ranked 44th in the Top-500 supercomputer list [20].

## 2.3. Target electronic structure

We use the software package ATOM [21] to peform the calculations of the quantities $E_{nl}$, $f_{nl}$ and $\sigma_{nl}$ entering Eq. (6). These results are presented in Table 1 and displayed graphically in Figure 3. In the table, we also list the binding energies returned by our TDSE code that utilizes the LHF potential. Except for the ground state energy, the binding energies $E_{nl}$ are close between the LHF and non-local HF calculations.

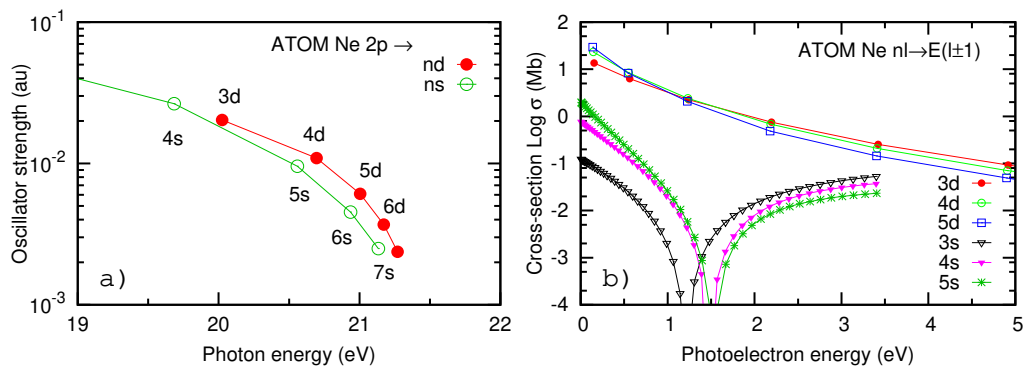

**Figure 3.** a) Oscillator strengths of the $2p \to nd$ and $2p \to ns$ transitions in Ne from the ATOM calculation. b) Partial photoionization cross-sections for the $nl \to E(l \pm 1)$ transitions.

**Table 1.** Binding energies $E_{nl}$ and oscillator strengths $f_{nl}$ of Ne in atomic units. The TDSE calculations with the LHF potential are compared with the ATOM calculations with the non-local HF potential. The experimental binding energies are from [22]

| $nl$ | Binding energy $E_{nl}$, au Expt. NIST | Theory TDSE | ATOM | $f_{nl}$, au |
|---|---|---|---|---|
| $2p$ | 0.792 | 0.788 | 0.850 | |
| $3s$ | 0.181 | 0.171 | 0.175 | 0.1505 |
| $4s$ | 0.066 | 0.067 | 0.068 | 0.0265 |
| $3d$ | 0.056 | 0.055 | 0.056 | 0.0203 |
| $4d$ | 0.031 | 0.031 | 0.031 | 0.0109 |
| $5d$ | 0.020 | 0.020 | 0.020 | 0.0061 |

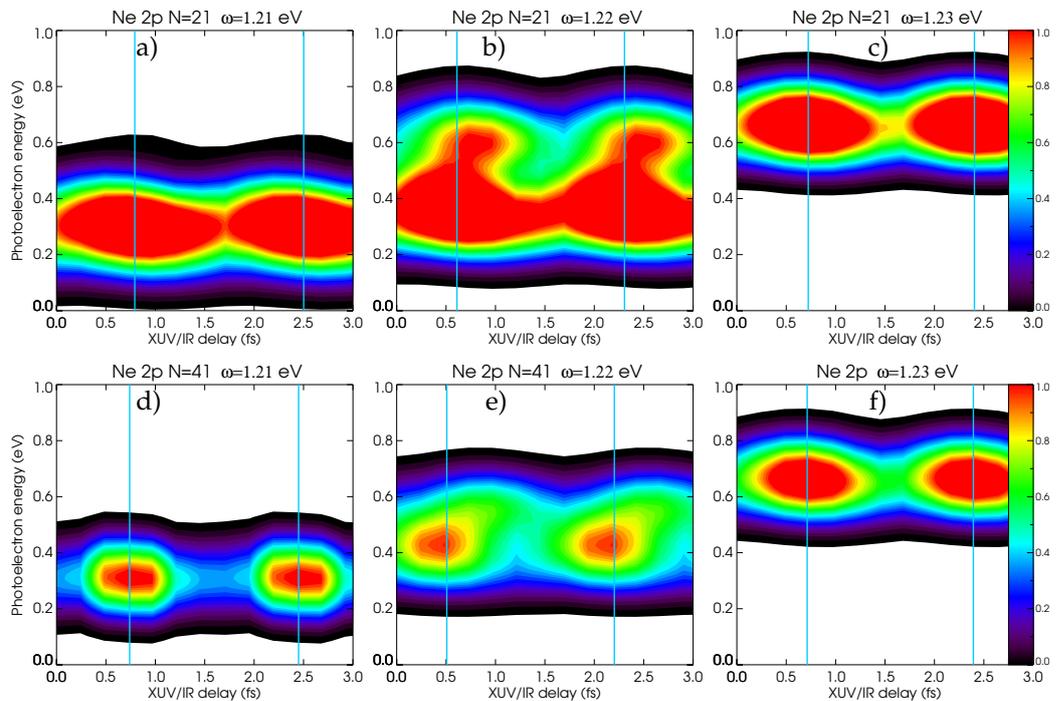

**Figure 4.** From left to right: RABBITT traces of the Ne atom at the photon frequencies $\omega = 1.21$, 1.22 and 1.23 eV. The number of the pulselets $N = 21$ in the top row corresponds to the spectral width FWHM=0.2 eV. In the bottom row, $N = 41$ and FWHM=0.1 eV. The thin blue lines mark the center of the SB18.

Figure 3a) shows that the oscillator strengths are close between the $ns$ and $nd$ bound states at comparable photon energies. However, Figure 3b) displays very clearly that the subsequent ionization process $nl + \omega \to E(l \pm 1)$ favours very strongly the $nd$ intermediate states and the $ns$ states can be safely ignored in uRABBITT. We see from the same Figure 3b) that the photoionization cross-secitons $\sigma_{nd}$ depend rather weakly on $n$ and it is the oscillator strength factor $f_{nd}^{1/2}$ that determines largely the resonant amplitude (6) in Ne.

## 3. Numerical results
### 3.1. Photolectron spectra

First we illustrate the effect of the XUV spectral width on the RABBITT photoelectron spectra. For this purpose, we analyze the RABBITT traces which are the stacks of the photoelectron spectra obtained at various time delay $\tau$ between the XUV and IR pulses. The RABBITT traces of the Ne atom at the IR frequencies of $\omega = 1.21$, 1.22 and 1.23 eV are shown in Figure 4 (from left to right). These traces illustrate the passage of the harmonic peak H17 over the 4$d$ bound state as is exhibited in Figure 2b). The RABBITT trace in Figure 4b) shows clearly the double hump structure resulting from the overlap of H17 both with the 4$d$ and 5$d$ bound states. The overlap with the single bound state 4$d$ or 5$d$ occurs at the photon energies $\omega = 1.21$ and 1.23 eV, respectively. These cases produce well-resolved single peak structures in Figure 4a) and c). The bottom row of panels in Figure 4(d-f) is analogous to the top row except the XUV spectral width is halved by increasing the number of the pulselets in the APT to $N = 41$. The 4*d* and 5*d* bound states are clearly resolved at all the photon energies.

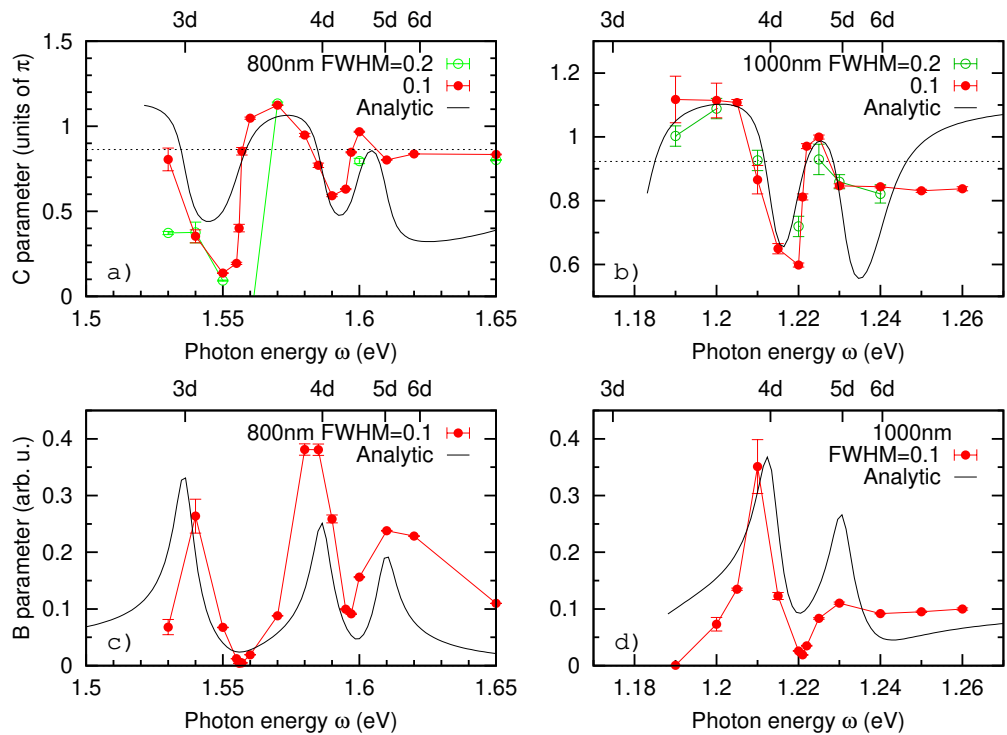

**Figure 5.** Top: the RABBITT phase *C* parameter as the function of the fundamental laser frequency in the 800 nm (a) and 1000 nm (b) wavelength ranges. The top horizontal axis marks the crossing of the harmonic peaks H13 (a) and H17 (b) with the discrete *nd* energy levels. Bottom: the RABBITT magnitude *B* parameter for the 800 nm (c) and 1000 nm (d) spectral ranges. Comparison with the analytic expressions (6) and (7) is made. The thin dotted line marks the non-resonant RABBITT *C* parameters which remain flat over the marked photon energy range.

### 3.2. RABBITT phase and magnitude parameters

The sideband intensities are integrated over the energy window $2q\omega \pm \text{FWHM}/2$ and their time dependence is fitted with Eq. (1). The resulting phase *C* parameters and the magnitude *B* parameters are plotted in the top and bottom rows of Figure 5, respectively. The left panels 5a) and c) correspond to the 800 wavelength range ($\omega = 1.53 - 1.65$ eV). The right panels 5b) and d) spans the photon energies in the 1000 wavelength range ($\omega = 1.19 - 1.26$ eV). The top horizontal axis marks the crossing of the submerged harmonic peak with a discrete energy level: $(2q-1)\omega_{nd} + E_i = E_{nd}$ where $(2q-1) = 13$ for 800 nm and 17 for 1000 nm, respectively. In Figure 5, the TDSE results with both FWHM=0.2 and 0.1 eV are plotted. The smaller XUV spectral width allows for a much more narrowly spaced photon energy points to be resolved.

The RABBIT phase *C* and magnitude *B* parameters are fitted with the analytic expressions (6) and (7). In the latter expression, the non-resonant terms are determined from the conventional SB16 for 800 nm and SB20 for 1000 nm. These terms remain flat over the considered photon energy range. Comparison of the TDSE and LOPT results is fair for both sets of parameters.

### 4. Summary and outlook

In the present work, we utilized the under-threshold RABBITT process to map the target electronic structure of the Ne atom. For this purpose, we run an extensive set of numerical TDSE simulations over the two spectral ranges near the central wavelengths of 800 and 1000 nm with a fine increment $\Delta\omega = 5$ meV. We made a comparison of our numerical results with predictions of the analytic LOPT expressions utilizing accurate bound state energies and oscillator strengths. These expressions predict a rapid variation of the RABBITT phase $C$ and the magnitude $B$ parameters when the submerged harmonic peak $H_{2q-1}$ overlaps with one of the $nd$ bound states. The companion $ns$ bound state series is found very weak in the Ne atom. Our numerical results agree rather well with the analytic formulas thus validating our approach. This means that the uRABBITT process can indeed be used for mapping the target atom electronic structure once the submerged harmonic peak sweeps across the series of the bound states. The uRABBITT technique can be used in various atomic targets and thus prove itself a useful and novel spectroscopic tool.

**Acknowledgments:**
The author gratefully acknowledges Giuseppe Sansone, Hamed Ahmadi and Matteo Moioli for many intense and stimulating discussions. Serguei Patchkovskii is acknowledged for placing his SCID TDSE code at the author's disposal. Resources of National Computational Infrastructure facility (NCI Australia) have been employed.

**Conflicts of Interest:** The author declares no known conflicts of interest

**Data Availability Statement:** The numerical data reported in the present work are available on request from the author